\documentclass[conference]{IEEEtran}
\IEEEoverridecommandlockouts
\usepackage{cite}
\usepackage{amsmath,amssymb,amsfonts}
\usepackage{algpseudocode}
\usepackage{algorithm}
\usepackage{graphicx}
\usepackage{textcomp}
\usepackage{xcolor}
\def\BibTeX{{\rm B\kern-.05em{\sc i\kern-.025em b}\kern-.08em
    T\kern-.1667em\lower.7ex\hbox{E}\kern-.125emX}}

\usepackage{tabularx}
\usepackage{seqsplit}
\usepackage{url}
\usepackage{multirow}
\usepackage{hyperref}

\begin{document}

\title{TTPXHunter: Actionable Threat Intelligence Extraction as TTPs from Finished Cyber Threat Reports\\
}

\author{Nanda Rani\textsuperscript{1}, Bikash Saha\textsuperscript{1}, Vikas Maurya\textsuperscript{1}, Sandeep Kumar Shukla\textsuperscript{1}\\
\textit{\textsuperscript{1}Department of Computer Science and Engineering, Indian Institute of Technology Kanpur,} \\ 
\textit{Kanpur, India}\\
\{nandarani,bikash,vikasmr,sandeeps\}@cse.iitk.ac.in}

\maketitle

\begin{abstract}
Understanding the modus operandi of adversaries aids organizations to employ efficient defensive strategies and share intelligence in the community. This knowledge is often present in unstructured natural language text within threat analysis reports. A translation tool is needed to interpret the modus operandi explained in the sentences of the threat report and convert it into a structured format. This research introduces a methodology named TTPXHunter for automated extraction of threat intelligence in terms of Tactics, Techniques, and Procedures (TTPs) from finished cyber threat reports. It leverages cyber domain-specific state-of-the-art natural language model to augment sentences for minority class TTPs and refine pinpointing the TTPs in threat analysis reports significantly. 
We create two datasets: an augmented sentence-TTP dataset of $39,296$ sentence samples and a $149$ real-world cyber threat intelligence report-to-TTP dataset. Further, we evaluate TTPXHunter on the augmented sentence and report datasets. The TTPXHunter achieves the highest performance of $92.42\%$ f1-score on the augmented dataset, and it also outperforms existing state-of-the-art TTP extraction method by achieving an f1-score of $97.09\%$ when evaluated over the report dataset.
TTPXHunter significantly improves cybersecurity threat intelligence by offering quick, actionable insights into attacker behaviors. This advancement automates threat intelligence analysis and provides a crucial tool for cybersecurity professionals to combat cyber threats.
\end{abstract}

\begin{IEEEkeywords}
Threat Intelligence, TTP Extraction, MITRE ATT\&CK, Natural Language Processing, Domain-specific Language model, TTP Classification, Cybersecurity
\end{IEEEkeywords}

\section{Introduction}
\label{sec:intro}

In the ever-evolving landscape of cybersecurity, Advanced Persistent Threats (APTs) represent a significant challenge to worldwide security. Countering APTs requires the development of sophisticated measures, which depend on the detailed extraction and analysis of threat intelligence related to APTs~\cite{Tang2022,rahman2023attackers,zhou2022cti}. It involves delving into the attacker's modus operandi in terms of Tactics, Techniques, and Procedures (TTPs)~\footnote{Categorized in the MITRE ATT\&CK Framework}~\cite{MITREATTACK} explained in the threat reports, blogs, bulletins released by security firms~\cite{bouwman2020different,daszczyszak2019ttp}. These reports, written in unstructured natural language, describe cyber adversaries' modus operandi. Converting this information into a machine-readable structured format improves threat intelligence efforts and is crucial for comprehending and mitigating potential threats~\cite{al2023mitre}.


Extracting TTPs from such reports is also crucial for recommending defensive mechanisms. However, this process often encounters challenges, including a lack of publicly available structured data, difficulties posed by polymorphic words, and the challenge of interpreting the contextual meanings of sentences present in the threat report and mapping them to MITRE ATT\&CK (Adversarial Tactics, Techniques, and Common Knowledge) TTP~\cite{Rani2023TTPHunter,rahman2022threat}. We combat these challenges in one of our previous works, which presents a TTP extraction tool named TTPHunter \cite{Rani2023TTPHunter}.
It can automatically identify and catalog TTPs with the f1-score of $0.88$~\cite{Rani2023TTPHunter}.
TTPHunter focused on the top 50 TTPs in the ATT\&CK framework because of the limited data availability for the remainder of the TTP class. The remainder of the TTPs are particularly those that are either newly emerging or less commonly used. This gap undermines the effectiveness of TTPHunter, as incomplete TTP intelligence can lead to a reactive rather than proactive security posture. Addressing this gap by advancing TTPHunter is crucial to ensure that threat intelligence is exhaustive and reflects the diverse adversarial tactics. 

\begin{table}[]
    \centering
    \caption{Notations and their Description}
    \label{tab:notations}
    \begin{tabularx}{\columnwidth}{|p{1cm}|X|}
    \hline
     \textbf{Notation} & \textbf{Description} \\
     \hline
     \hline
     $S_i$ & $i^{th}$ sentence present in the report \\
     \hline
     $w_i$ & $i^{th}$ word in the sentence \\
     \hline
     $n$ & No. of sentences present in the report \\
     \hline
     $Aug\_S$ & List of augmented sentences for the sentence $S$ \\
     \hline
     $\theta$ & Threshold to choose relevant sentences \\
     \hline
     $S_{embed}$ & Embedding vector for sentence $S$\\
     \hline
     $Top\_5$ & Top-5 selected words to create augmented sentence \\
     \hline
     $f(.)$ & Embedding function\\
     \hline
     $\mathcal{M}$ & Fine-tuned classifier\\
     \hline
     $v$ & Feature vector\\
     \hline
     $t_i$ & Predicted TTP class for $i^{th}$ sentence in the report\\
     \hline
     $\Theta$ & Threshold to filter irrelevant sentences from threat report \\
     \hline
     $SCORE$ & Similarity score between augmented and original sentence\\
     \hline
     $\hat{y}$ & True label multi-hot vector for given threat report\\
     \hline
     $y$ & Predicted label multi-hot vector for given threat report\\
     \hline
     $HL$ & Hamming Loss\\
     \hline
     $\mathcal{R}$ & A threat report which consists of $n$ sentences $(S)$ \\
     \hline
     $\mathcal{T}$ & Set of unique TTPs present in the database \\
     \hline
     $m$ & No. of total unique TTPs present in the database \\
     \hline
     $\hat{\mathcal{T}}$ & Predicted set of TTPs \\
     \hline
     $d$ & No. of words present in the sentence\\
     \hline
\end{tabularx}
\end{table}

Therefore, this study presents TTPXHunter, an extended version of TTPHunter~\cite{Rani2023TTPHunter}, which meticulously refines and expands to recognize an impressive array of $193$ TTPs.
We address the limited data problem of TTPHunter by introducing an advanced data augmentation method. This method meticulously preserves contextual integrity and enriches our training dataset with additional examples for emerging and less commonly used TTPs. Moreover, we also employ a domain-specific language model that is finely tuned to grasp the nuanced, context-driven meanings within this domain and enhance both the augmentation process and the TTP classification. Further, we also enable TTPXHunter to convert the extracted TTPs into the machine-readable format, i.e., STIX, which facilitates the automated exchange, easier analysis, and integration across different security tools and platforms~\cite{barnum2012standardizing,briliyant2021towards}. 
This advancement significantly broadens the scope of threat intelligence gleaned from threat reports and offers deeper insights into the TTPs employed by cyber adversaries in APT campaigns. By extending TTPHunter's extraction capabilities, this research contributes to the critical task of threat intelligence gathering, providing security analysts and practitioners with a more comprehensive toolset for identifying, understanding, and countering APTs. The notations and their descriptions used to explain the methodology are listed in Table.~\ref{tab:notations}.


TTPXHunter applies to a threat report denoted as $\mathcal{R}$ and extracts set of TTPs denoted as $\hat{\mathcal{T}}$ explained within the report, where $\hat{\mathcal{T}} \subseteq \mathcal{T}$ and $\mathcal{T} = \{t_1,t_2,t_3,\ldots,t_m\}$ which denotes a set of target labels having $m$ number of TTPs.
The methodology initiates with the transformation of sentences segmented from threat report, denoted as $\mathcal{R} = \{S_1, S_2,\ldots, S_n\}$; $n$ represents the number of sentences in the report, into a high-dimensional feature space, i.e., $768$-dimension. This transformation, achieved through a cyber domain-specific embedding function $f(.)$, maps each segmented sentence $S_i$ to a $768$-dimension feature vector $v_i$ as
\begin{equation*}
    v_i = \{f(S_i) \text{ } ; \quad \forall S_i \in \mathcal{R}\}
\end{equation*}
The embedding function encapsulates cybersecurity language's rich semantic and syntactic properties and the stage for nuanced inference. TTPXHunter exhibits an inferential capability to predict a specific TTP  for each segmented sentence based on its embedded feature vector:
\begin{equation*}
    t_i = \mathcal{M}(v_i) ; \quad t_i \in \mathcal{T}
\end{equation*}
Where $t_i$ represents the predicted TTP class for the feature vector $v_i$, and $\mathcal{M}$ embodies the fine-tuned classification model that aids in understanding the complex relationship between the features of the segmented sentence and the TTP class.
Further, the TTP extraction from the threat report $\mathcal{R}$ is performed as:
\begin{equation*}
    \hat{\mathcal{T}} = \bigcup_{i=1}^{n} t_i
\end{equation*}

Our key contributions to this research are the following:
\begin{itemize}
    \item We introduce TTPXHunter\footnote{We plan to include the source code link in the camera-ready version.}, an extended version of TTPHunter \cite{Rani2023TTPHunter}, which extracts TTPs from threat intelligence reports using the SecureBERT \cite{Aghaei2023SecureBERT} language model. We fine-tune the model on our prepared augmented sentence-TTP dataset and convert the extracted TTPs into structured and machine-readable STIX format.
    \item We introduce a data augmentation method that utilizes the cyber-domain-specific language model. This approach creates sentences related to TTPs by preserving their contextual meaning while preparing new and varied sentences.
    \item By leveraging the presented data augmentation method, we build an augmented sentence-TTP dataset using the TTPHunter dataset prepared from the MITRE ATT\&CK knowledgebase. We build the augmented dataset using the MLM (Masked Language Model) feature of the contextual natural language model, and it extends samples from $10,906$ sentences to $39,296$ sentences covering $193$ ATT\&CK TTPs. 
    \item We perform an extensive evaluation to measure the efficiency of TTPXHunter over two different types of datasets: The augmented sentence dataset and the Threat report dataset. We manually label $149$ real-word threat reports for the report dataset.
\end{itemize}

The remainder of the paper is structured as follows: Section \ref{sec:relwork} discusses the current literature, Section \ref{sec:background} presents the required background, Section \ref{sec:methodology} discusses presented TTPXHunter method, Section \ref{sec:ExpEval} demonstrate the experiment done and evaluate the obtained results, Section \ref{sec:limitationfuture} presents the limitation in proposed method along with it is possible future direction and Section \ref{sec:conclusion} concludes this research contribution.





\section{Related Work}
\label{sec:relwork}

\begin{table*}[!ht]
    \centering
    \caption{TTP Extraction Methods Comparison}
    \label{tab:TTP_LR}
    \begin{tabularx}{\textwidth}{|X|X|p{4.5cm}|X|X|X|X|X|}
\hline
\textbf{Research Work} & \textbf{Year} & \textbf{Extraction Technique} & \textbf{Sentence Context-aware} & \textbf{Identify Relevant Text} & \textbf{STIX Support} & \textbf{Domain-specific Capability} & \textbf{Range of MITRE ATT\&CK TTPs} \\
\hline
\hline
Husari et al.~\cite{Husari2017Ttpdrill}  & $2017$ & TF-IDF, Ontology-based, and improved BM25 similarity rank & $\times$ & $\checkmark$ & $\checkmark$ & $\checkmark$ & All TTPs\\
\hline
Legoy et al.~\cite{legoy2020automated} - rcATT & $2020$ & TF-IDF and ML Models (KNN, Decision Tree, Random Forest, and Extra Tree) & $\times$  & $\times$ & $\checkmark$ & $\times$ & All TTPs\\
\hline 
Li et al.~\cite{Li2022AttacKG} - AttacKG & $2022$ & Entity-based dependency graph and Graph-alignment algorithm & $\times$ & $\checkmark$ & $\times$ & $\checkmark$ & All TTPs\\
\hline
Rani et al.~\cite{Rani2023TTPHunter} - TTPHunter. & $2023$ & BERT/RoBERTa followed by Linear Classifier & $\checkmark$ & $\checkmark$ & $\times$ & $\times$ & 50 most frequently used TTPs only\\
\hline
Alam et al.~\cite{alam2023looking} - LADDER & $2023$ & Extract attack phases using a sequence tagging model and map these patterns to TTP using cosine similarity & $\checkmark$ & $\checkmark$ & $\times$ & $\times$ & All TTPs \\
\hline
TRAM~\cite{Tram} & $2023$ & SciBERT followed by Linear Classifier & $\checkmark$ & $\times$ & $\times$ & $\times$ & 50 most frequently used TTPs only \\
\hline
\textbf{TTPXHunter (Proposed)} & $2024$ & \textbf{Domain-specific SecureBERT with Linear Classifier} & $\checkmark$ & $\checkmark$ & $\checkmark$ & $\checkmark$ &  \textbf{All TTPs}\\
\hline

\end{tabularx}
\end{table*}


Research on TTP extraction from threat intelligence reports is widely based on ontology, graphs, and keyword-phrase matching methods~\cite{rahman2022threat,Rani2023TTPHunter}.

Initially, Husari et al. \cite{Husari2017Ttpdrill} present TTPDrill, which extracts threat actions based on ontology and matches TTP's knowledge base with extracted threat action using the BM25 matching technique. Legoy et al. \cite{legoy2020automated} present rcATT, a classification tool based on Machine Learning (ML) algorithms. They use Term Frequency–Inverse Document Frequency (TF-IDF) to prepare the dataset and perform multi-class classification for target labels as TTP. The rcATT is based only on sentence keywords, as TF-IDF ignores the sentence's word sequence and context. The model may fail when synonym words are present and polymorphic nature words (the same words have a different meaning in a different context). Li. et al. \cite{Li2022AttacKG} introduce AttacKG, a graph template matching technique. The method obtains IOCs and constructs entity-based dependency graphs for every TTP present in ATT\&CK by leveraging descriptions on the MITRE website and matching them with the TTP's templates prepared from MITRE website data. To match the templates, they use the graph alignment algorithm. AttacKG struggles to capture techniques identified by adjectives (properties of IOCs) present in the sentences rather than verbs (threat actions), such as masqueraded identity and obfuscated malware.
Alam et al. ~\cite{alam2023looking} introduce LADDER, a framework designed to enhance cyber threat intelligence (CTI) by extracting and analyzing attack patterns from CTI reports, addressing the limitations of traditional CTI that focuses on static indicators like IP addresses. LADDER contains a subpart named TTPClassifier, which is structured into three key steps: identifying sentences with attack pattern descriptions using a binary classification model, pinpointing and extracting these attack phrases with a sequence tagging model, and finally, mapping these patterns to TTP IDs via cosine similarity method. 
To deal with the polymorphic nature of words and leverage contextual information in the report sentences, Rani et al. \cite{Rani2023TTPHunter} present TTPHunter. This tool leverages the language models BERT and RoBERTa to understand the context of the sentences present in the dataset and map it to the correct TTP ID present in the dataset. Due to a dearth of sentence datasets for many TTPs, TTPHunter is trained for only $50$ sets of TTPs and has yet to map the full spectrum of TTPs present in the MITRE knowledge base.
After that, MITRE also introduced a TTP extraction tool named TRAM (Threat Report ATT\&CK Mapping)~\cite{Tram}. It is based on the scientific BERT model, named SciBERT~\cite{beltagy2019scibert}, a fine-tuned BERT model on a collection of scientific reports.

In this paper, we extend our recent work TTPHunter \cite{Rani2023TTPHunter} to extract TTPs present in the collected threat report. The TTPHunter is fine-tuned on traditional BERT and RoBERTa models, whereas we fine-tune cyber-domain-specific language models to identify TTPs. Aghaei et al. \cite{Aghaei2023SecureBERT} show that the cyber-domain-specific language model can perform better than the traditional model trained on general English sentences. Domain-specific words such as Windows and registry have different meanings regarding general and cybersecurity usage. In addition, we solve the limited dataset problem pointed out by \cite{Rani2023TTPHunter} using the data augmentation method. Our proposed method can extract the full spectrum of TTPs in the MITRE ATT\&CK knowledgebase. A comparison of our proposed model TTPXHunter with the literature is shown in Table \ref{tab:TTP_LR}.







\section{Background}
\label{sec:background}

This section presents the details of every framework, tool, and method used to get insight into their background, which is required to understand the methodology.

\begin{figure*}
    \centering
    \includegraphics[height=8cm, width=\textwidth]{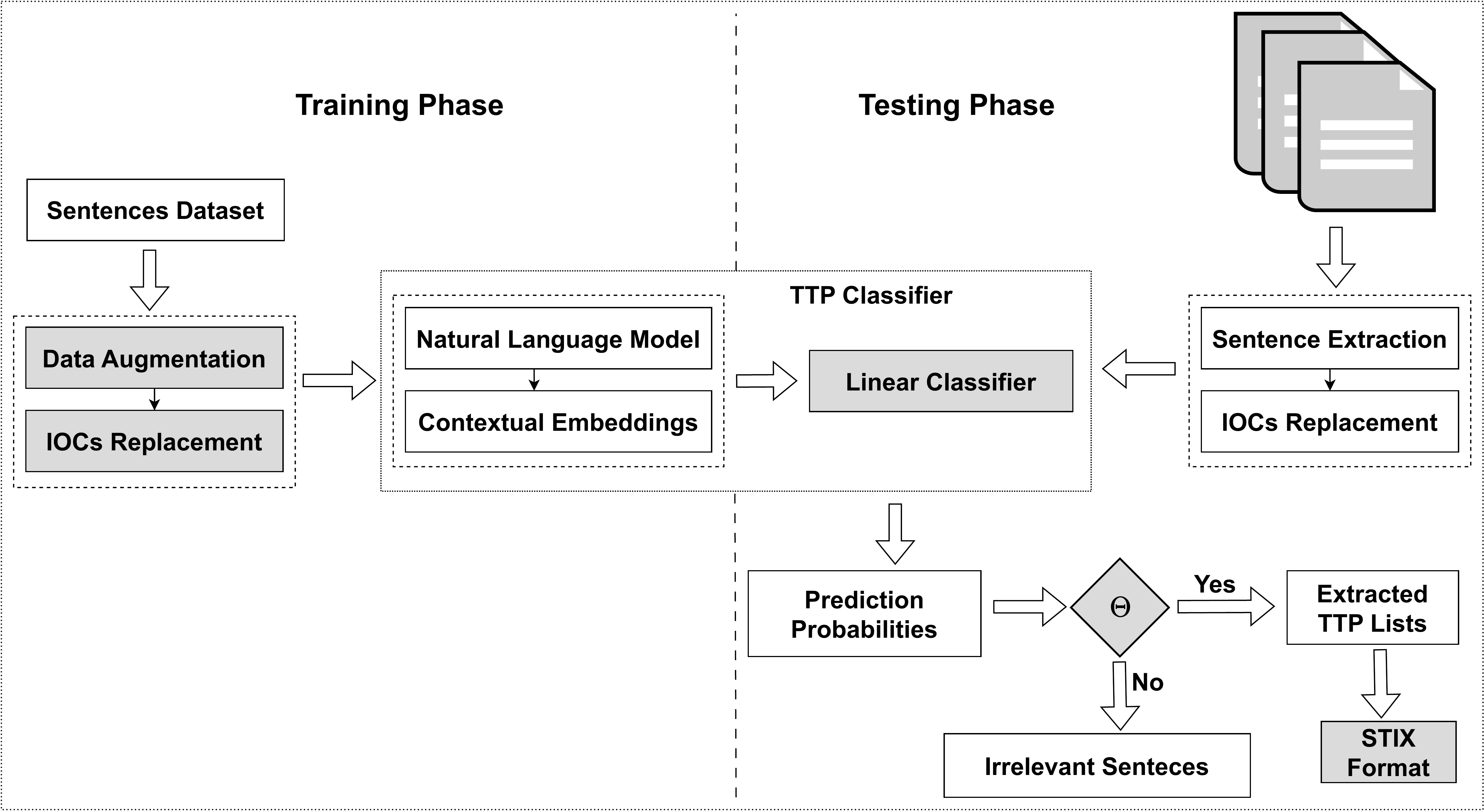}
    \caption{TTPXHunter Architecture}
    \label{fig:TTPXHunter}
\end{figure*}

\subsection{MITRE ATT\&CK Framework}
\label{subsec:MITREframework}
The MITRE ATT\&CK (Adversarial Tactics, Techniques, and Common Knowledge) Framework \cite{MITREATTACK} is a globally acknowledged comprehensive knowledge base designed to understand cyber threats and adversary behaviors across different stages of a cyber attack. It introduces a standardized lexicon and framework for the classification and detailed description of attackers' tactics, techniques, and procedures (TTPs). The ATT\&CK framework comprises three primary components: Tactics, Techniques, and Procedures. Tactics represent the objectives or goals adversaries aim to achieve throughout an attack, encompassing $14$ distinct categories such as Reconnaissance, Resource Development, Initial Access, Execution, and others~\cite{MITREATTACK}. Techniques detail the specific methods adversaries employ to fulfill these tactics, each with a unique identifier and a comprehensive description. Examples include phishing, password spraying, remote code execution, and PowerShell exploitation. Procedures delve into the detailed execution of techniques by attackers to achieve their tactical objectives, illustrating how specific methods are applied in practice, such as deploying a spearphishing email with malicious attachments to achieve initial access. Beyond cataloging techniques, the ATT\&CK framework maps real-world attack scenarios to known techniques that enable organizations to understand better how adversaries navigate through the stages of an attack. This knowledge fosters the development of proactive defenses and offensive security practices, enhances incident response efforts, and enriches threat intelligence~\cite{daszczyszak2019ttp,ajmal2021offensive}.

\subsection{BERT Language Model}
\label{subsec:BERT}
The natural language model plays a vital role in cyber threat intelligence, particularly in extracting threat intelligence from natural language cybersecurity texts. Among the popular language models, BERT (Bidirectional Encoder Representations from Transformers) \cite{Vaswani2017Attention} holds significant importance in tasks such as extracting threat intelligence, classifying threat data, NER (Name Entity Recognition), detecting spam and phishing attacks \cite{Aghaei2023SecureBERT,Tikhomirov2020Using,Rifat2022BERT}. BERT comprises a 12-layer stacked encoder part of the transformer, which generates contextualized embeddings for natural language inputs.
The training of the BERT model involves two key components: Masked Language Model (MLM) and Next Sentence Prediction (NSP). In the MLM, random words within a sentence are masked, and the model learns to predict the masked word based on the context of surrounding words present in the sentence. The NSP focuses on understanding the context and meaning of a sentence by predicting the likelihood of the next sentence in a given pair of sentences. Both tasks demonstrate the BERT model's ability to grasp the contextual understanding of sentences and capture the relationships between words within a sentence. The BERT model's proficiency in understanding sentence context and word relationships makes it valuable for various downstream tasks in cyber threat intelligence.

\subsection{TTPHunter}
\label{subsec:TTPHunter}
Our recent tool, TTPHunter~\cite{Rani2023TTPHunter}, leverages the power of the language model BERT to extract threat intelligence in the form of TTPs from natural language threat report texts. The authors of TTPHunter fine-tune a pre-trained BERT model using a sentence-TTP dataset collected from the MITRE knowledge base to let the model understand the context of TTPs present in the MITRE ATT\&CK matrix and map a given sentence to relevant TTPs. One noteworthy feature of TTPHunter is its ability to discern the significance of sentences. Rather than indiscriminately mapping all sentences from a threat report, TTPHunter selectively identifies and considers only those sentences that truly explain TTPs, disregarding irrelevant information. To achieve this, we implement a filtering mechanism on the model's predicted probabilities. Our experiments discovered that sentences with a probability higher than $0.64$ are relevant, while those below the threshold were considered irrelevant. We adopt the same threshold for identifying relevant sentences by TTPXHunter.
\section{Proposed Methodology: TTPXHunter}
\label{sec:methodology}

\begin{algorithm*}
\caption{Data Augmentation Algorithm}\label{alg:augmentation}
\textbf{Input:} $S \gets [w_1, w_2, \dots, w_d]$ \Comment{Input Sentence consisting $d-$words} \\
\textbf{Output:} $Aug\_S$ \Comment{List of augmented sentences for the input sentence}
\begin{algorithmic}[1]
    \State $Aug\_S \gets [\;]$
    \For{$i=1$ to $d$}
        \State $S' \gets S$
        \State $S'[i] \gets <mask>$ \Comment{Mask the $i^{th}$ word in sentence S}
        \State $Predicted\_words \gets \Call{SecureBERTMLM}{S'}$ \Comment{Predict probable word using MLM}
        \State $Top\_5 \gets \Call{Select\_top\_5}{Predicted\_words}$ \Comment{Select top-5 probable words based on their probabilities}
        \For{$word$ in Top\_5}
            \State $S'[i] \gets word$ \Comment{Augmented sentence obtained by replacing masked token}
            \State $SCORE \gets \Call{Cosine\_Similarity}{S,S'}$
            \If{$SCORE \geq \theta$}
                \State $Aug\_S \gets Aug\_S \cup S'$ \Comment{Append augmented sentence to output list}
            \EndIf
        \EndFor
    \EndFor
    \State \textbf{Return} $Aug\_S$
    \vspace{0.2cm}
    \Function{Cosine\_Similarity}{$S,S'$}
        \State $S_{embed} \gets \Call{Sentence\_Transformer}{S}$
        \State $S'_{embed} \gets \Call{Sentence\_Transformer}{S'}$
        \State \textbf{Return} $cosine(S_{embed}, S'_{embed})$
    \EndFunction
\end{algorithmic}
\end{algorithm*}

In this study, we introduce TTPXHunter, an extended version of TTPHunter, designed to overcome the challenges of limited data for less commonly encountered TTPs and improve the performance of TTPHunter. It incorporates a novel methodology encompassing data augmentation and domain-specific language models to enhance the performance of TTP classification. The notations used to explain the methodology are listed in Table.~\ref{tab:notations}.

TTPXHunter follows the idea used in TTPHunter of leveraging the natural language model to extract TTPs from threat reports. TTPHunter is based on the model, which is fine-tuned on general English sentences.
The model trained on general sentences can mislead the contextual embedding for domain-specific words having different contexts comparatively. Cyber domain-specific words such as Windows and Registry have entirely different meanings in the cyber domain. Hence, we observe the need for a TTP extraction tool based on domain-specific language models to provide more accurate contextual embeddings.
To fill this gap of domain-specific knowledge, we leverage the SecureBERT~\cite{Aghaei2023SecureBERT} to fine-tune the proposed TTPXHunter and map sentences to the relevant TTPs they represent. TTPXHunter incorporates a filtration mechanism to identify and exclude irrelevant sentences from the TTP extraction results obtained from the threat reports. The filtering mechanism ensures the model extracts the most pertinent and meaningful TTPs for further analysis and discards unrelated sentences. The overview and architecture of TTPXHunter are shown in Fig.~\ref{fig:TTPXHunter}.

The TTPHunter~\cite{Rani2023TTPHunter} is limited to $50$ prominent TTPs only out of $193$ TTPs due to the limited sentence in the database for the remainder of the TTP class, which results in incomplete or limited 
threat intelligence. Enhancing its extraction capabilities is crucial for developing comprehensive and proactive security strategies by leveraging the full spectrum of threat intelligence. Therefore, we address this problem by presenting a data augmentation method in TTPXHunter. This method creates more samples for the minority TTP class. 

\subsection{Contextual Data Augmentation}
\label{subsec:DataAug}

\begin{table*}[!ht]
    \centering
    \caption{IOC Replacement}
    \begin{tabularx}{\textwidth}{|c|c|X|}
        \hline
        \textbf{Base Name} & \textbf{Example Pattern} & \textbf{Regex} \\
        \hline
        \hline
        Registry & \texttt{HKEY\_LOCAL\_MACHINE\textbackslash{}XXX\textbackslash{}XXX\textbackslash{}XX\textbackslash{}Run} & \texttt{\seqsplit{(HKEY\_LOCAL\_MACHINE|HKEY\_CURRENT\_USER|HKEY\_CLASSES\_ROOT|HKEY\_USERS|HKEY\_CURRENT\_CONFIG)\textbackslash{}\textbackslash{}(?:[\textasciicircum\textbackslash{}\textbackslash{}]+\textbackslash{}\textbackslash{})*[\textasciicircum\textbackslash{}\textbackslash{}]+}} \\
        \hline
        Email & \texttt{example@example.com} & \texttt{\seqsplit{[a-zA-Z0-9.\_{}\%\textbackslash+-]+@[a-zA-Z0-9.-]+\textbackslash.[a-zA-Z]\{2,\}}} \\
        \hline
        IP Address & \texttt{00.00.00.00} & \texttt{\seqsplit{(\textbackslash{}d\{1,3\}\textbackslash{}.)\{3\}\textbackslash{}d\{1,3\}}} \\
      \hline
      Domain & \texttt{www.domain.com} & \texttt{\seqsplit{[a-zA-Z0-9.-]+\textbackslash{}.[a-zA-Z]\{2,\}}} \\
      \hline
      File Path & \texttt{C:\textbackslash{}XXX\textbackslash{}yy} & \texttt{\seqsplit{([a-zA-Z]:\textbackslash{}\textbackslash{})?(?:[a-zA-Z0-9\_-]+\textbackslash{}\textbackslash{})*[a-zA-Z0-9\_-]+\textbackslash{}.[a-zA-Z0-9\_]+}} \\
      \hline
      File Name & \texttt{file.exe} & \texttt{\seqsplit{[a-zA-Z0-9\_-]+\textbackslash{}.[a-zA-Z0-9\_]+}} \\
      \hline
      CVE & \texttt{CVE-YYYY-XXXX} & \texttt{\seqsplit{CVE-\textbackslash{}d\{4\}-\textbackslash{}d\{4,\}}} \\
      \hline
\end{tabularx}
    \label{tab:IOC}
\end{table*}

\begin{figure}[!ht]
    \centering
    \includegraphics[width=0.5\textwidth]{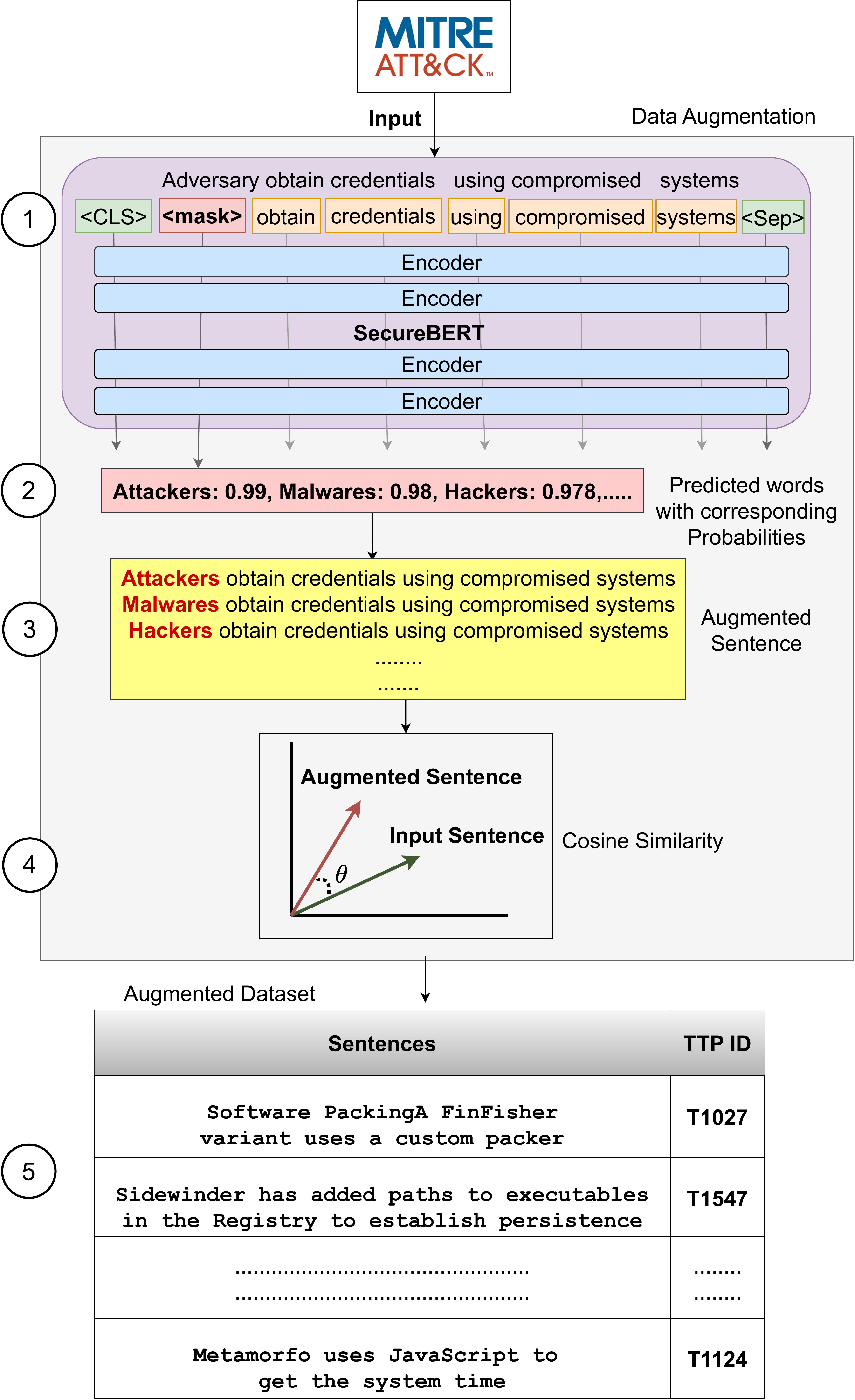}
    \caption{Data Augmentation Steps }
    \label{fig:DataAug}
\end{figure}

We leverage the MLM (Masked Language Model) capability of the natural language model to expand the dataset and address the limited samples problem of TTPHunter for the remainder of the TTP class. For MLM, we employ the domain-specific language model called SecureBERT \cite{Aghaei2023SecureBERT}. In this process, we mask words in each sentence with a special token $<mask>$ and employ SecureBERT to predict the masked word using its MLM capability. SecureBERT provides a list of candidate words and their corresponding probabilities, which maintain the contextual meaning when replacing the masked word.

We can see at step {\large \textcircled{\normalsize 1}} in Fig. \ref{fig:DataAug}, an example of input sentence as \texttt{"Adversary obtained credentials using compromised systems"}. In this case, if we mask the word \texttt{"Adversary"}, then the model predicts possible words that preserve the contextual meaning (step {\large \textcircled{\normalsize 2}} in Fig. \ref{fig:DataAug}). The probabilities associated with each word indicate the confidence level of the predicted word.
We select the top $5$ words from this list and generate five new sentences for each input sentence (step {\large \textcircled{\normalsize 3}} in Fig. \ref{fig:DataAug}). The sentence's meaning may deviate after replacing the masked word with the predicted word. 
Therefore, we employ cosine similarity to compare each newly generated sentence with the original sentence (step {\large \textcircled{\normalsize 4}} in Fig. \ref{fig:DataAug}) and select the sentence with the highest similarity index. To compute the cosine similarity, we generate contextual embedding for both the original and generated sentences using Sentence Transformer \cite{Reimers2019Sentence}.
The similarity score ranges between $0$ (not similar) to $1$ (Exactly similar), and we set a threshold $(\theta)$ of $0.975$. We retain only those sentences having similarity scores greater than $\theta$.
The sentence augmentation algorithm is outlined in Algorithm \ref{alg:augmentation}, which we employ to construct the augmented sentence-TTP dataset. More detailed information regarding the distribution of the augmented sentence-TTP dataset is present in the section \ref{subsec:Dataset}.

\subsection{Prepossessing \& Fine-Tuning}
The sentences present in the dataset consist of irrelevant structures, which we fix during the pre-processing method and let the data add more value to the dataset. First, we remove the citation references from these sentences, which refer to past attack campaigns or threat reports. Further, we find various IOCs (Indicator of Compromise) patterns present in the sentences, which potentially obscure the contextual understanding of the sentences. For example, IP and domain addresses, file paths, CVE IDs, emails, and registry paths. We implement an IOC replacement method to overcome the obstruction these patterns impose in sentences. This method uses regular expressions (regex) to substitute IOC patterns with their respective base names. For example, a sentence like 
\texttt{"}Upon execution, the malware contacts the C2 server at \texttt{attacker-example.com}, 
drops an executable \texttt{payload.exe} at
\texttt{\seqsplit{C:$\backslash$Users$\backslash$Default$\backslash$AppData$\backslash$Roaming}}, 
and creates an autorun entry in \texttt{\seqsplit{HKEY\_LOCAL\_MACHINE$\backslash$Software$\backslash$Microsoft$\backslash$Windows$\backslash$CurrentVersion$\backslash$Run}"} get transformed to \texttt{"}Upon execution, the malware contacts the C2 server at \texttt{domain}, 
drops an executable \texttt{file} at \texttt{file path}, and creates an autorun entry in \texttt{registry}\texttt{"}.
This method lets the model add more contextual information rather than hindering the meaning, which helps better TTP identification. Table \ref{tab:IOC} details the considered IOCs example, their corresponding base name, and the use of regex pattern. The processed sentences get passed further for fine-tuning.

\begin{figure*}
    \centering
    \includegraphics[width=\textwidth]{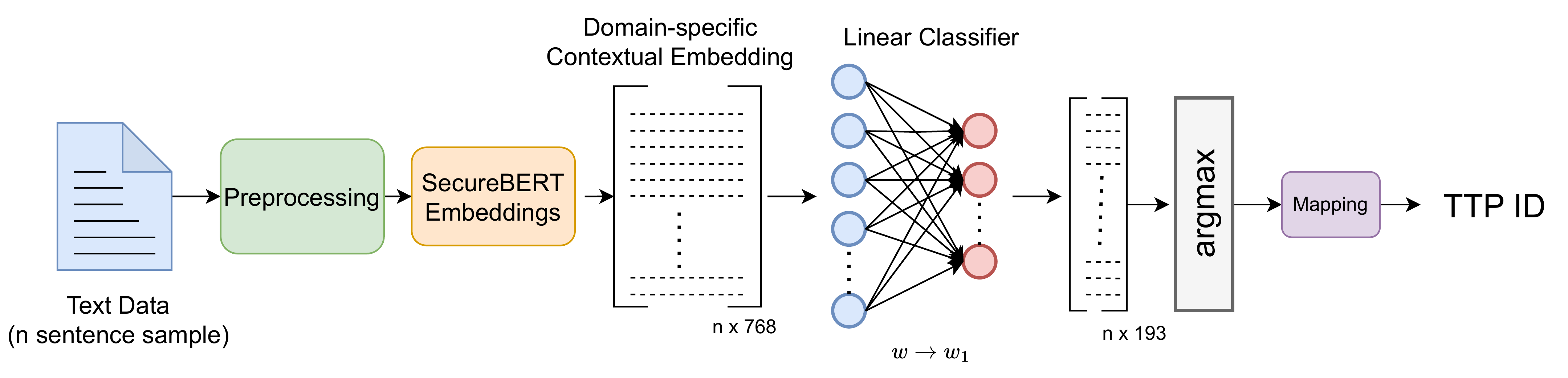}
    \caption{Fine tuning-steps}
    \label{fig:finetune}
\end{figure*}

The finetuning process is similar to the TTPHunter's finetuning. As TTPHunter extracts sentence embedding from the traditional BERT model and passes it to a linear classifier for classification, we employ a similar method in TTPXHunter. It first extracts domain-specific embedding using the SecureBERT language model and passes embedding to a linear classifier for TTP classification. The finetuning process is meticulously orchestrated to optimize performance on the processed sentence dataset. We initiate the process by configuring essential parameters, setting the learning rate to $1e-5$ to ensure subtle adjustments to the model, choosing a batch size of $64$ to balance computational efficiency with training effectiveness, and training it for $10$ epochs. We leverage the Hugging Face's Transformers library~\cite{HuggTrans} for its robust support of transformer models. Sentences are tokenized using SecureBERT's tokenizer, aligning with its pre-trained understanding of language structure, and inputs are standardized to a maximum length of $256$ tokens. We finetune our model to let it understand the context for ATT\&CK TTP, which further enhances the classification capability.
The model's finetuning is shown in Fig.~\ref{fig:finetune}.

\subsection{TTP Extraction}
TTPXHunter takes finished threat reports as input and breaks the report contents into a list of individual sentences. Then, each sentence is processed and passed to the TTPXHunter's finetuning architecture for classification. The system classifies each sentence to identify TTPs. Finally, it aggregates these TTPs and creates a list of TTPs extracted from the analyzed report.
A threat report usually contains many irrelevant sentences that do not reflect TTPs. However, our classification module is a closed-world solution (the input will surely be classified into at least one target class). As a result, irrelevant sentences can also be classified as one of the TTPs in the target class, which can lead to huge false positives. To reduce this effect, we filter irrelevant sentences, similar to TTPHunter. We employ a threshold mechanism in the classification module to filter irrelevant sentences and only map sentences that explain TTPs, i.e., a relevant sentence. We filter irrelevant sentences based on classification confidence score. We fix a threshold $(\Theta)$ and filter the sentences if the classifier's confidence score is below the threshold.
We follow the same threshold value experimentally obtained in TTPHunter, i.e., $0.644$.

Once the linear classifier extracts the list of TTPs present in the threat report, TTPXHunter converts the list to the Structured Threat Information eXpression (STIX)~\cite{barnum2012standardizing} format, significantly enhances cyber threat intelligence operations. It supports detailed analysis and correlation of threats by providing a rich, structured representation of cyber threat information. This structured format facilitates automated processing and threat response, which increases operational efficiency. Moreover, the consistency and standardization offered by STIX improve communication within and between organizations to ensure a common understanding of cyber threats. 



\section{Experiments and Results}
\label{sec:ExpEval}
This section presents details about the prepared dataset and the experiments. We also discuss the results obtained and chosen performance measures.

\subsection{Dataset}
\label{subsec:Dataset}
We compute model performance on our test dataset through sentence-to-TTP mapping. However, in the real world, the threat data is in the form of reports containing a set of sentences mixed with relevant and irrelevant sentences. Therefore, we also evaluate the model's performance report-wise.
As a result, we prepared two datasets to measure the efficiency of TTPXHunter and the current literature: $1)$ The augmented Sentence-TTP dataset and $2)$ The Report-TTP dataset.
\subsubsection{Augmented Sentence-TTP Dataset}
\label{subsubsec:SenteceDataset}
To prepare the sentence-TTP dataset, we consider the TTPHunter dataset as a base dataset, prepared using MITRE ATT\&CK knowledgebase \cite{Rani2023TTPHunter}. The base dataset consists of two columns: sentences and their corresponding TTP ID. The dataset consists of $10,906$ sentences over $193$ TTPs. We use our proposed data augmentation algorithm, explained in Algorithm \ref{alg:augmentation}, and extend the dataset to $39,296$ sentences distributed over $193$ TTP classes. The TTP ID $T1059$ (Command and Scripting Interpreter) consists of the highest number of sentences as $800$, and TTP ID $T1127$ (Trusted Developer Utilities Proxy Execution) consists of the lowest number of samples as $3$. On average, the number of samples in our dataset is $203$. The distribution of data samples is present in Appendix \ref{apxsubsec:SenteceDataDist} and a glimpse of the dataset is shown at step {\large \textcircled{\normalsize 5}} in Fig \ref{fig:DataAug}.

\subsubsection{Report-TTP Dataset}
\label{subsubsec:DocumentDataset}
Generally, threat data is present in the form of threat analysis reports. Therefore, We evaluate TTPXHunter on a document dataset, which demonstrates the performance of TTPXHunter in filtering irrelevant sentences from threat reports. It also tells us how efficiently TTPXHunter can extract threat intelligence from threat reports. We manually collect $149$ threat reports published by various prominent security firms, and we manually label the set of TTPs present in each report to prepare the ground truth. The document-TTP dataset contains two columns: threat report and list of TTP present in the corresponding threat report.


\subsection{Evaluation}
\label{subsec:Exp}
We evaluate TTPXHunter using various performance metrics and compare its performance with several current studies. As our dataset is imbalanced, shown in Appendix Fig. \ref{fig:sentenceTTPdist}, we consider macro-averaged precision, recall, and f1-score~\cite{scikit-learn}. This approach ensures balanced evaluation across all classes~\cite{grandini2020metrics}. By giving equal importance to each TTP class, these metrics prevent the majority class's dominance from overshadowing the minority class's performance. It promotes the development of effective and fair models across different TTP classes and is essential for nuanced TTP classification tasks.

\subsubsection{Augmented Sentence-based Evaluation}
We finetune TTPXHunter on the prepared augmented dataset and compare the other two BERT-based models, i.e., TRAM~\cite{Tram} and TTPHunter~\cite{Rani2023TTPHunter} present in the literature.
We divide the prepared augmented dataset into train and test sets with an $80:20$ ratio.


\par
\begin{figure}
    \centering
    \includegraphics[width=\columnwidth]{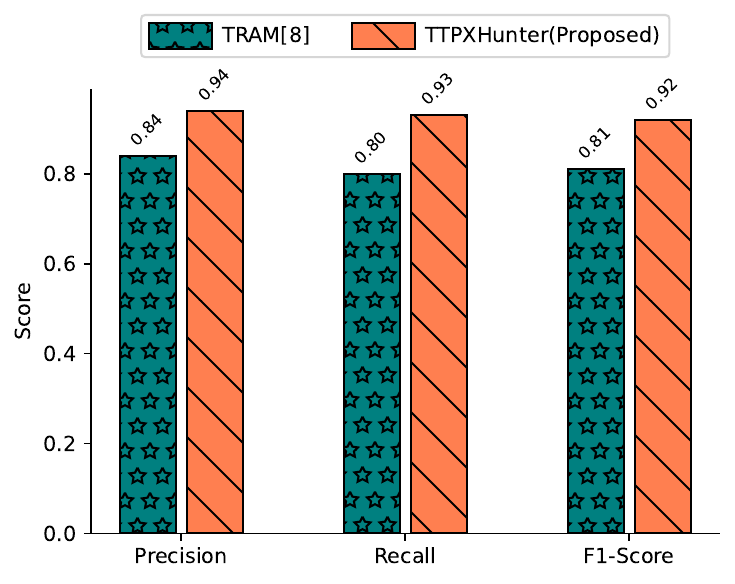}
    \caption{Performance Comparison between TRAM~\cite{Tram} and TTPXHunter over Augmented Dataset}
    \label{fig:ttpxhuntertram}
\end{figure}

\textbf{TTPXHunter vs TRAM : }
We fine-tune the TTPXHunter on the train set and evaluate its performance using the chosen performance metrics. Further, we finetune the literature TRAM~\cite{Tram} on the same dataset and evaluate its performance. By employing TRAM on the augmented dataset, we extend the capability of TRAM from the 50 most frequently used TTPs to the full spectrum of TTPs. As a result, it gives common ground for comparing TTPXHunter and TRAM. The result obtained by both methods and their comparison is shown in Fig.~\ref{fig:ttpxhuntertram}. As we can see, the TTPXHunter outperforms the TRAM, which reflects the difference between contextual embedding of general scientific BERT and domain-specific BERT embeddings. This result reflects that the domain-specific language model provides a better contextual understanding of embedding than the language model trained on general scientific terms.


\par
\textbf{TTPXHunter vs TTPHunter : }
We assess the performance of our proposed TTPXHunter alongside state-of-the-art TTPHunter.
Our proposed extraction method, TTPXHunter, performs better than state-of-the-art TTPHunter. TTPXHunter's superiority is due to using a cyber-domain-specific finetuned language model. Sentences containing domain-specific terms, such as "window" and "registry," introduce a distinct context that differs from general English. This distinction allows our method, based on the domain-specific language model, to capture and interpret the contextual meaning more accurately than traditional models.
In addition, TTPXHunter can identify the range of $193$ TTPs with $0.92$ f1-score, whereas TTPHunter is limited to only $50$ TTPs. The improvement in the result and the capability to identify all ranges of TTPs make TTPXHunter superior to TTPHunter.

\begin{figure}
    \centering
    \includegraphics[width=\columnwidth]{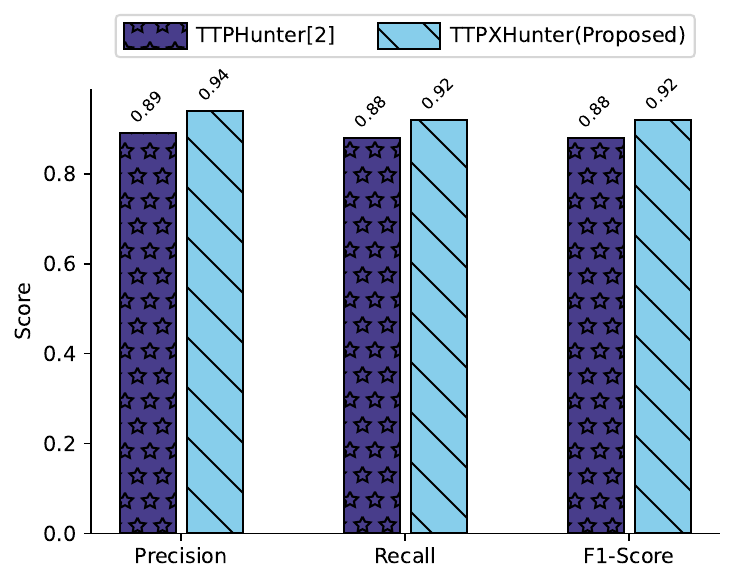}
    \caption{Performance Comparison between TTPHunter~\cite{Rani2023TTPHunter} and TTPXHunter over TTPHunter's 50-TTP set}
    \label{fig:ttpxhuntttphunt}
\end{figure}


Further, to understand the efficiency of the data augmentation method and compare the performance of TTPXHunter and TTPHunter on the same base, we evaluate TTPXHunter on the ground of TTPHunter. We only selected 50 TTP sets for which we developed TTPHunter and evaluated both. The obtained result is shown in Fig.~\ref{fig:ttpxhuntttphunt}. As we can see, TTPXHunter outperforms TTPHunter on the 50-TTP set ground of TTPHunter. It reflects that augmenting more samples for a 50-TTP set and employing a domain-specific language model enables the classifier to better understand the context of TTP.








\subsubsection{Report-based Evaluation}
\label{subsubsec:RepoExp}

\begin{table*}[!ht]
    \centering
    \caption{Comparison with State-of-the-art methods over Report Dataset}
    \begin{tabular}{|c|c|c|c|c|}
    \hline
    \textbf{Model} & \textbf{Precision (\%)} & \textbf{Recall (\%)} & \textbf{F1-score (\%)} & \textbf{Hamming Loss} \\
    \hline
    \hline
     ATTACKG~\cite{Li2022AttacKG} & $88.58$ & $95.22$ & $88.52$ & $0.14$ \\
     \hline
     rcATT~\cite{legoy2020automated} & $30.47$ & $44.03$ & $30.56$ & $0.64$ \\
     \hline
     Ladder~\cite{alam2023looking} & $92.97$ & $95.73$ & $93.90$ & $0.10$ \\
     \hline
     TRAM~\cite{Tram} & $94.54$ & $94.33$ & $93.49$ & $0.10$ \\
     \hline
     \textbf{TTPXHunter (Proposed)} & $\textbf{97.38}$ & $\textbf{96.15}$ & $\textbf{97.09}$ & $\textbf{0.05}$\\
     \hline
\end{tabular}
    \label{tab:ResultComparisonMacro}
\end{table*}

In the real world, we have threat reports in the form of natural language rather than sentence-wise datasets, and these reports contain information along with TTP-related sentences. So, we evaluate the TTPXHunter on the report dataset, which contains each sample as threat report sentences and a list of TTP explained in the report. Extracting TTPs from threat reports is a multi-label problem because a list of TTP classes is expected as output for any given sample, i.e., threat report in this case. The evaluation of such classification also requires careful consideration because of multi-label classification.
\par
\textbf{Evaluation Metrics:} In the multi-label problem, the prediction vector appears as a multi-hot vector rather than a one-hot vector in a multi-class problem. In the multi-label case, there may be a situation where not all expected TTP classes were predicted; instead, a subset of them is correctly predicted. However, the prediction may be wrong because the whole multi-hot vector does not match. For example, If the true label set contains $\{T1, T2, T3\}$ and the predicted label is $\{T2, T3\}$, then it may be considered to be a mismatch even though $T2$ and $T3$ are correctly classified. So, relying on accuracy may not be a good choice for multi-label problems~\cite{saha2023malxcap}; instead, we consider hamming loss as a performance metric to deal with such a scenario. The hamming loss measures the error rate label-wise~\cite{de2009tutorial,zhang2013review,saha2023malxcap}. It calculates the ratio of incorrect labels to all labels. For given $k$ threat reports, the hamming loss is defined as:
\begin{equation}
\label{eq:hamingloss}
    \text{HL} = \frac{\sum_{i=1}^{k}[y_i \oplus \hat{y}_i]}{k}
\end{equation}
Where, $y_i \text{ and } \hat{y}_i$ are multi-hot predicted labels and true label for $i^{th}$ instance, respectively.  The $\oplus$ represents element-wise exclusive OR (XOR) operation. The low hamming loss represents that models make minimal wrong predictions.
\par
Further, we also evaluate macro precision, recall, and f1-score by leveraging a multi-label confusion matrix package from sklearn~\cite{scikit-learn}. Then, we calculate true positive, false positive, and false negative for each class and calculate these performance metrics. Further, we calculate the macro average between all classes to get macro-averaged performance metrics for all chosen measures, i.e., precision, recall, and f1-score. We prefer the macro-average method to ignore biases towards the majority class and provide equal weight to all classes.
\\
\\
We consider four state-of-the-art methods, i.e., \cite{legoy2020automated,Li2022AttacKG,alam2023looking,Tram} for comparison against TTPXHunter based on these metrics over the report dataset. This comparison aims to understand the effectiveness of TTPXHunter over state-of-the-art for TTP extraction from finished threat reports. These methods provide the list of TTPs extracted from the given threat report and the model confidence score for each. Out of all extracted TTPs, only relevant TTPs are selected based on the threshold mechanism decided by each method. We evaluate the state-of-the-art method's performance based on the threshold value given in their respective papers. For TTPXHunter, we obtain the same threshold experimentally chosen for TTPHunter, i.e., $0.644$.
The results obtained from all implemented methods on our report dataset are shown in Table~\ref{tab:ResultComparisonMacro}. It demonstrates that TTPXHunter outperforms all implemented methods across all chosen metrics, i.e., the lowest hamming loss and the highest other performance metrics. It achieves the highest f1-score of $97.07\%$, whereas out of all state-of-the-art methods, LADDER~\cite{alam2023looking} performs better than other state-of-the-art methods and achieves $2^{nd}$ highest performance of $93.09\%$ F1-score. 
This performance gain over state-of-the-art methods demonstrates the efficiency of the TTPXHunter, and we plan to make it open for the benefit of the community.

As this experiment involves $193$ target TTP classes, it is challenging to visualize the class-wise performance of the employed models. Therefore, we follow a different way to assess the class-wise efficiency of the employed models. We count the number of TTP classes whose chosen performance metrics lie within a range. We employ a range interval of $0.1$, i.e., $10\%$, to calculate the number of TTP classes whose score falls into the range.
\\

\begin{figure*}
    \centering
    \includegraphics[width=\textwidth]{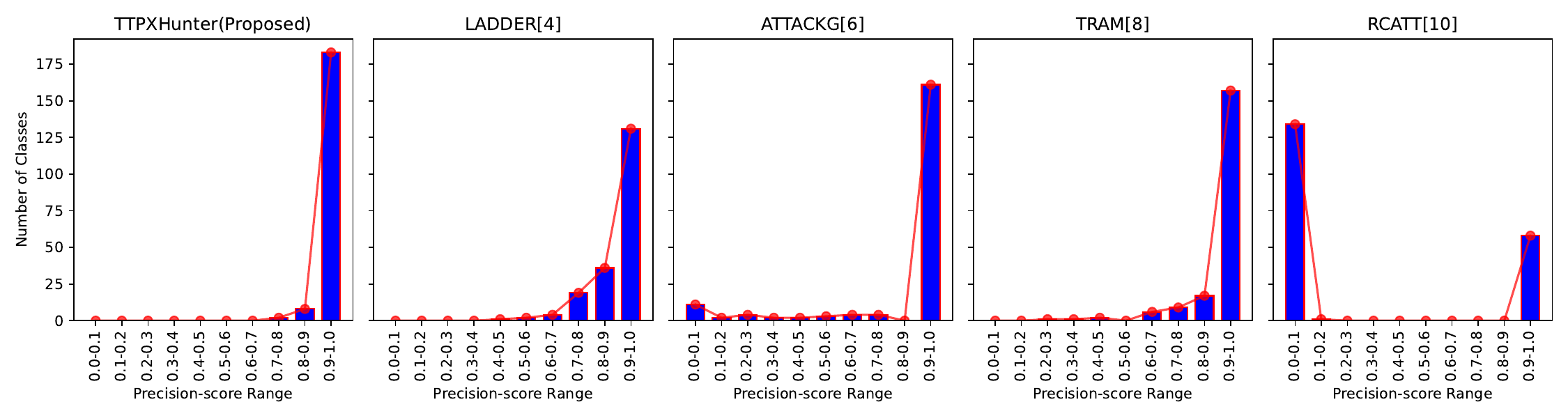}
    \caption{Class-wise Precision Score Comparison with State-of-the-art Methods}
    \label{fig:ClasswisePrecision}
\end{figure*}

\begin{figure*}
    \centering
    \includegraphics[width=\textwidth]{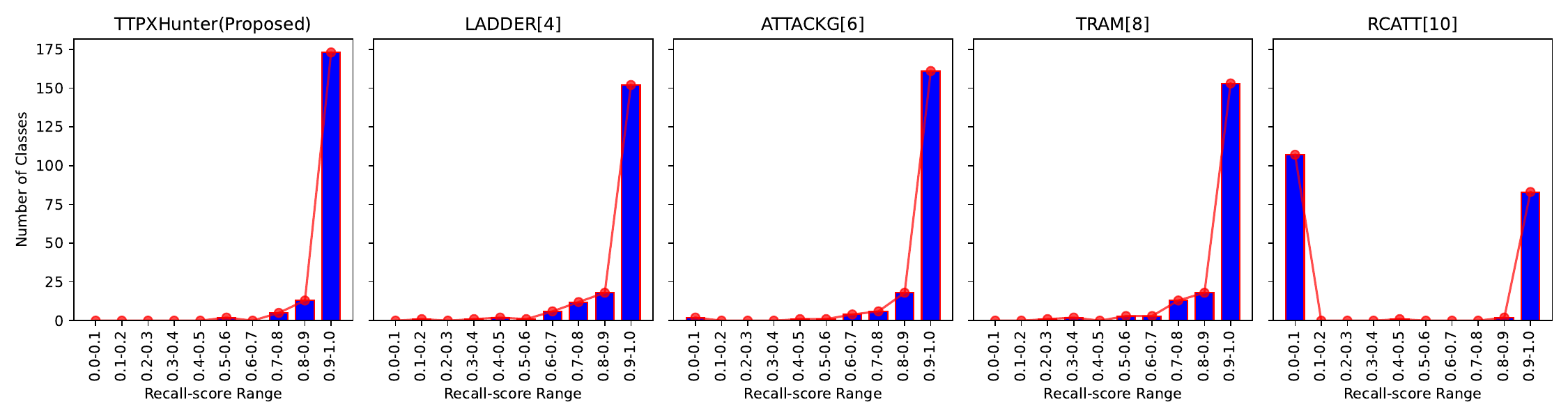}
    \caption{Class-wise Recall Score Comparison with State-of-the-art Methods}
    \label{fig:ClasswiseRecall}
\end{figure*}

\begin{figure*}
    \centering
    \includegraphics[width=\textwidth]{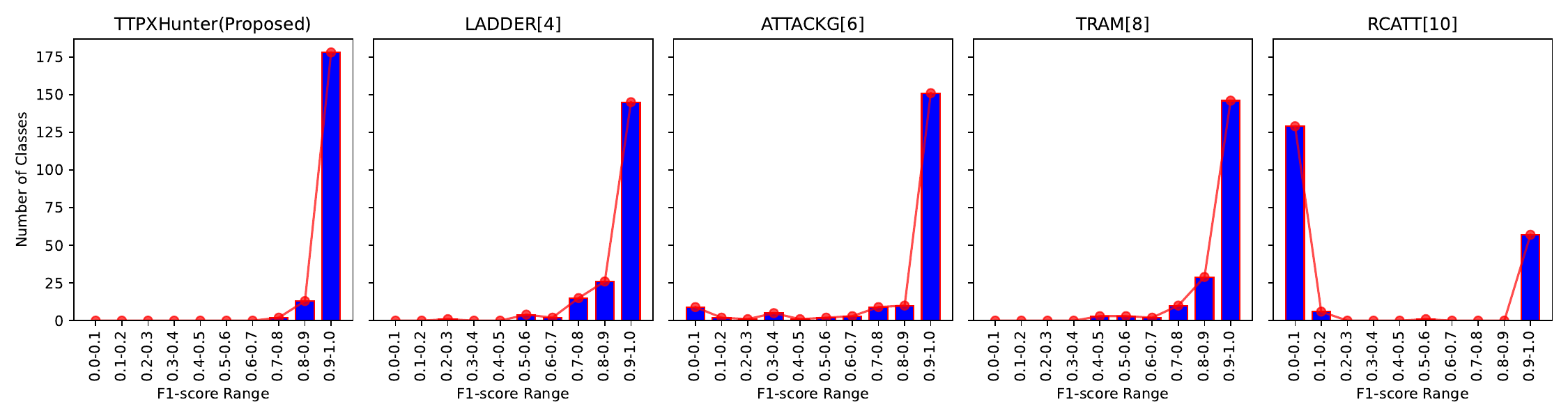}
    \caption{Class-wise F1 Score Comparison with State-of-the-art Methods}
    \label{fig:Classwisef1}
\end{figure*}

We perform this calculation across all five methods and three chosen performance metrics. The obtained results are present in Figures~\ref{fig:ClasswisePrecision}, \ref{fig:ClasswiseRecall}, and \ref{fig:Classwisef1}. The observation reveals that most TTP classes analyzed by rcATT fall within the $0-0.10$ range, contributing to its overall lower performance. This performance is due to the reliance on the TF-IDF method to transform sentences into vectors. TF-IDF is a numerical statistic that reflects how important a word is to a document in a collection or corpus, balancing its frequent appearance within a document against its commonness across all documents~\cite{salton1968computer,sparck1972statistical,salton1975vector}.
This method lacks the ability to understand the context and semantic relationships between words, making it unable to grasp the overall meaning of sentences~\cite{selva2021review,patil2023survey}. ATTACKG, conversely, exhibits a lower score within the $0-0.10$ range for certain TTP classes, which adversely affects its overall performance. However, TRAM shows a minimum performance range of $0.3-0.4$ for TTP classes, which indicates better performance than rcATT and ATTACKG. LADDER ensures a performance score of at least $0.4-0.5$ for a TTP class, which positions it ahead of the aforementioned methods, including rcATT, ATTACKG, and TRAM. Our proposed model, TTPXHunter, assures a minimum score of $0.6-0.7$ for TTP classes, with the majority exhibiting scores between $0.9-1.0$, which underscores TTPXHunter's significant advantage over the other methods. It demonstrates the effectiveness of domain-specific models for domain-specific downstream tasks.


\section{Limitations \& Future Directions}
\label{sec:limitationfuture}

In addition to the advantages of TTPXHunter, like threat intelligence extraction, threat profiling, and sharing, it has some limitations that one should be careful about while using for actionable insights. The MITRE ATT\&CK framework is not a kind of fixed knowledge base. Instead, MITRE threat researchers are continuously updating it. One may need to finetune the model again if new TTPs are added to get the newer TTPs predicted. While finetuning on newer TTPs may require using our proposed data augmentation method to create new augmented sentences specific to newer TTPs. Therefore, we plan to make it public so one can adapt our method for any new upcoming versions of the ATT\&CK framework.

Further, the TTPXHunter contains a one-to-one classifier model, which maps a given sentence to a single TTP. Like a single sentence, a sentence containing one-to-many mapping may explain more than one TTP. For example, the sentence is \texttt{"}The attacker gained initial access through a phishing email and obtained persistence via run registry modification\texttt{"}. TTPXHunter maps this sentence to T1566 (Phishing) or T1037 (Boot or Logon Initialization Scripts) in such a scenario. Extending the capability of TTPXHunter to identify such one-to-many mapping can help us improve the model's performance. We plan to take up this challenge to improve the efficiency of TTPXHunter in the future.
\section{Conclusion}
\label{sec:conclusion}

This research demonstrates the efficiency of domain-specific language models in extracting threat intelligence in terms of TTPs, which can share the attack patterns and accelerate the threat response and detection mechanisms. The tool TTPXHunter extends the TTPHunter's capability over multiple dimensions, such as expanding to the full spectrum of TTP extraction and improving efficiency by leveraging domain-specific language models. We evaluate the efficiency of TTPXHunter over the prepared augmented sentence-TTP dataset and report-TTP dataset. On the augmented dataset, TTPXHunter outperforms both BERT-variant models, i.e., TRAM and TTPHunter. TTPXHunter also outperforms the state-of-the-art TTP extraction methods by achieving the highest F1 score and lowest hamming loss. TTPXHunter's performance over the report dataset demonstrates the model's efficiency in capturing relevant sentences from threat reports and correctly classifying them to the TTP class. The conversion of extracted TTPs to STIX makes integrating threat intelligence into security operations easier. TTPXHunter aids in improving the threat analyst's capability to share intelligence, analyze threats, understand the modus operandi of sophisticated threat actors, and emulate their behavior for red teaming. Therefore, TTPXHunter can support various cybersecurity teams, including red, blue, and purple teams in an organization.

\appendix

\subsection{Sentence-TTP Dataset Distribution}
\label{apxsubsec:SenteceDataDist}
\begin{figure*}[h]
  \centering
  \begin{minipage}{\textwidth}
    \rotatebox{90}{%
      \begin{minipage}{\textheight}
        \centering
        \includegraphics[width=\textwidth]{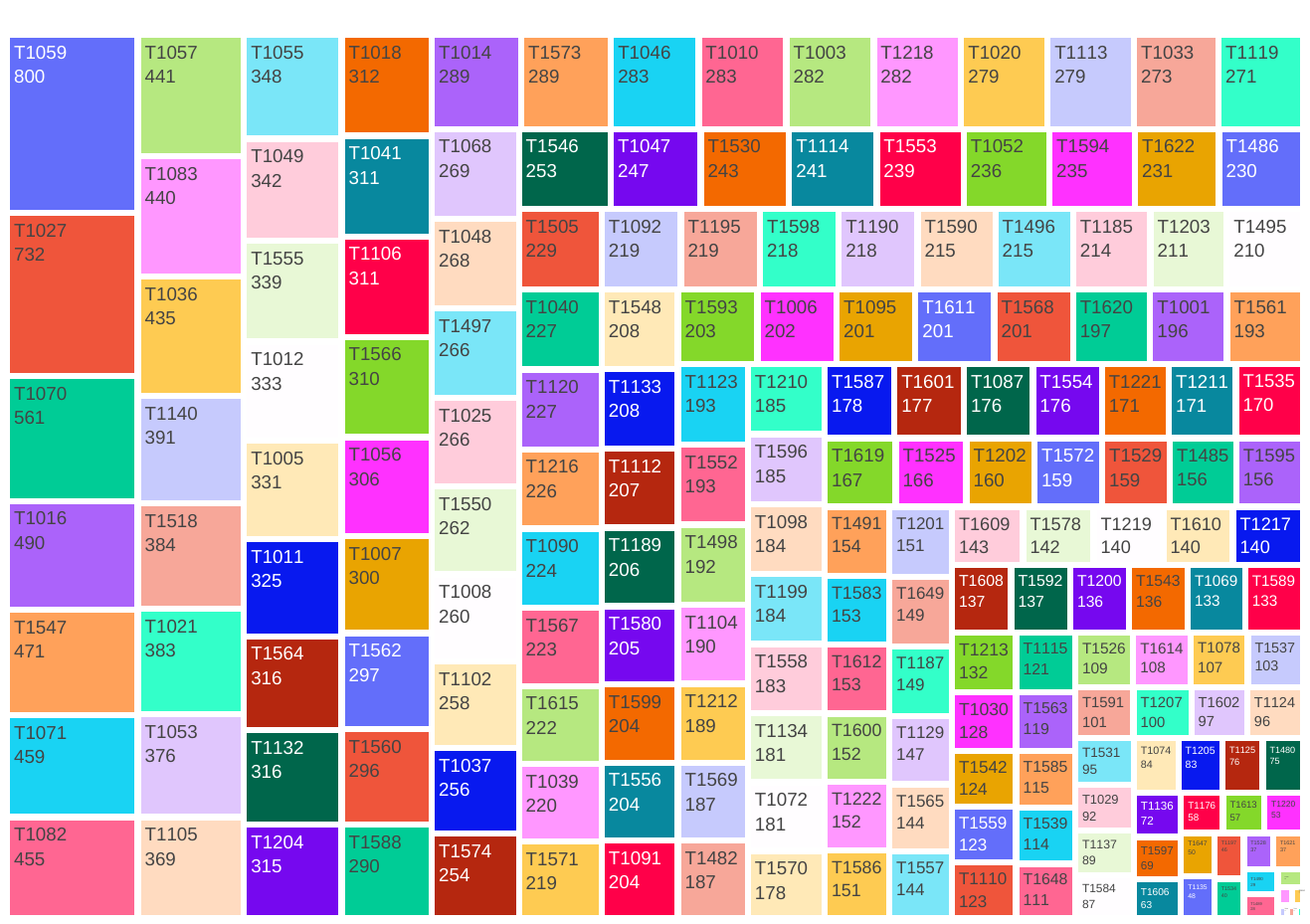}
        \caption{Sentence-TTP Dataset Data Distribution over 193-TTP Class}
        \label{fig:sentenceTTPdist}
      \end{minipage}
    }
  \end{minipage}
\end{figure*}

This section plots the Sentence-TTP dataset distributed over the $193$-TTP classes. The sentence-TTP dataset contains sentences as data sample and their corresponding TTP ID. The number of samples in each TTP class is shown in Fig \ref{fig:sentenceTTPdist}. In this figure, each tile represents a block for each TTP class, and contains TTP ID and their corresponding number of samples present in the dataset.
~


\end{document}